\documentclass[useAMS,usenatbib]{mn2e}

\pdfoutput=1

\usepackage{float}
\usepackage{enumerate}
\usepackage{multirow}
\usepackage{array}
\usepackage{comment}

\usepackage{amssymb,amsfonts,amsmath,graphicx,multirow}

\title[Minimum Atmospheric Height (MAH)]{A simple, quantitative method to infer 
the minimum atmospheric height of small exoplanets\thanks{Code available at
http://goo.gl/aZD9a}}
\author[Kipping, Spiegel \& Sasselov]{David M. Kipping$^{1,2}$\thanks{E-mail:
dkipping@cfa.harvard.edu}, David S. Spiegel$^{3}$ \& Dimitar D. Sasselov$^{1}$  \\
$^{1}$Harvard-Smithsonian Center for Astrophysics, 60, Garden Street, Cambridge, MA 02138, USA \\
$^{2}$Carl Sagan Fellow \\
$^{3}$Astrophysics Department, Institute for Advanced Study, Princeton, NJ 08540, USA}
\begin{document}

\date{Accepted 2013 June 10. Received 2013 June 3; in original form 2013 May 3}

\pagerange{\pageref{firstpage}--\pageref{lastpage}} \pubyear{2013}

\maketitle

\label{firstpage}

\begin{abstract}

Amongst the many hundreds of transiting planet candidates discovered by the
\emph{Kepler} mission, one finds a large number of candidates with sizes between 
that of the Earth and Neptune. The composition of these worlds is not 
immediately obvious with no Solar system analogue to draw upon and there exists
some ambiguity as to whether a given candidate is a rocky super-Earth or
a gas-enveloped mini-Neptune. The potential scientific value and observability 
of the atmospheres of these two classes of worlds varies significantly and given
the sheer number of candidates in this size-range, there is evidently a need for
a quick, simple metric to rank whether the planets have an extended atmosphere
or not. In this work, we propose a way to calculate the ``minimum atmospheric
height'' ($R_{\mathrm{MAH}}$) using only a planet's radius and mass as inputs.
We assume and exploit the boundary condition that the bulk composition of a 
solid/liquid super-Earth cannot be composed of a material lighter than that of 
water. Mass-radius loci above a pure-water composition planet correspond to
$R_{\mathrm{MAH}}>0$. The statistical confidence of a planet maintaining an
extended atmosphere can be therefore easily calculated to provide a simple 
ranking of target planets for follow-up observations. We also discuss how this 
metric can be useful in the interpretation of the spectra of observed planetary 
atmospheres.

\end{abstract}


\section{Introduction}
\label{sec:intro}

In recent years, the characterization of exoplanetary atmospheres has become
of both increasing interest and feasibility thanks to the large number of
confirmed exoplanets and a burgeoning number of instruments capable of 
measuring the associated effects \citep{tinetti:2009,seager:2010}. Transit
spectroscopy and emission spectroscopy have emerged as the most widely used
techniques to this end, with observers constraining the chemical composition of
the atmospheres of several worlds to date (e.g. \citealt{charbonneau:2002,
tinetti:2007,bean:2011,sing:2012}). The very large number of exoplanets, more 
than 850 at the time of writing (www.exoplanet.eu; \citealt{schneider:2011}), 
and the resource-intensive nature of the required observations to measure 
exoplanetary atmospheres (e.g. \citealt{agol:2010,fraine:2013}) mean that 
observers are forced to select only the most promising exoplanets for further 
study. This selection is typically based on the simple premise of focusing on 
those exoplanets where one should expect to detect the largest signal-to-noise
ratio, e.g., bright target stars and large-radius planets. 

Increasingly, the study of small exoplanets ($R_P\lesssim3$\,$R_{\oplus}$) is 
becoming feasible, thanks to the discovery of transiting planets around small 
M-dwarf stars \citep{charbonneau:2009} and the use of improved instrumentation 
\citep{berta:2012}. The study of the atmospheres of such small exoplanets will 
likely become increasingly prevalent as observers seek to push down to more 
telluric-like planets combined with the windfall of low-radius planets detected
by \emph{Kepler} \citep{batalha:2013,fressin:2013,dong:2012}. Careful target 
selection for atmospheric characterization of these small exoplanets will 
therefore become crucial for future planned missions e.g. \emph{EChO}
\citep{tinetti:2012}.

One challenge with studying exoplanets with radii 
$R_P\simeq 1.5$-$3$\,$R_{\oplus}$ is that such worlds straddle the 
boundary between rocky, terrestrial worlds (``super-Earths'') and small gaseous
planets (``mini-Neptunes''). Naturally this classification has a significant
impact on the prior probability of a detectable atmosphere and the 
interpretation of a spectrum. In traditional core accretion theory, 
runaway gas accretion is predicted for planetary embryos exceeding 
$\sim10$\,$M_{\oplus}$ \citep{pollack:1996} leading to large 
hydrogen/helium-dominated atmospheres. Assuming a rocky/icy core, the 
$\sim10$\,$M_{\oplus}$ transition corresponds to $\sim2$\,$R_{\oplus}$ 
\citep{valencia:2006}. For this reason, \emph{Kepler}'s recent discovery 
\citep{batalha:2013} of a large population of small exoplanets with radii 
close to this transition ($R_P\simeq 1.5$-$3$\,$R_{\oplus}$), along with
the apparently smooth distribution across this regime, was not generally 
expected.

In any case, an inescapable conclusion is that distinguishing whether a
specific exoplanet is a mini-Neptune or a rocky super-Earth is not possible 
with a radius measurement alone. Consequently, a catalogue of exoplanetary 
radii is insufficient for selecting targets for follow-up atmospheric
characterization in the regime of $R_P\simeq 1.5$-$3$\,$R_{\oplus}$.

Another challenge is that it has become evident that many degeneracies exist 
in the process of spectral retrieval \citep{benneke:2012}, particularly
salient when the spectrum is relatively flat as in the case of GJ 1214b 
for example \citep{bean:2011,berta:2012}. Here, a flat spectrum can be
considered consistent with either a low mean molecular weight atmosphere 
with clouds or a high mean molecular weight atmosphere yielding a low scale 
height. This invites the inclusion of additional priors to constrain the 
various models.

Despite the described pains of interpreting super-Earths/mini-Neptunes, there
does exist at least one major advantage. Specifically, a considerable range of
the expected internal pressures of such worlds are achievable in the laboratory
(unlike Jovian worlds), meaning that the phase diagrams of the constituent 
molecules in the planet's core can be empirically calibrated. Accessing these 
pressures (up to 100\,GPa) has only recently been achievable and was
utilized in the recent revised mass-radius relationship presented by
\citet{zeng:2013}. Such mass-radius relationships impose at least two hard
boundary conditions: i) a maximum $M_P$-$R_P$ contour for a pure-iron
planet and ii) a minimum $M_P$-$R_P$ contour for a pure-water planet. In
principle, it is not expected to find an object with a mass exceeding that
of boundary condition i), except for exotic states of matter found in the
core of stars/stellar remnants. In contrast, it is possible to find an
object with a mass below that of boundary condition ii) but such an object
must maintain an atmosphere\footnote{One could of course imagine pathological 
compositions such as pure lead that would violate boundary condition i) or pure 
CO$_2$, which might violate boundary condition ii), but such counterexamples 
seem implausible.}. Several effects such as intense irradiation and dissolved 
gas may affect the robustness of boundary condition ii) and we will address 
these later in \S\ref{sec:discussion}. 

Here, we show how this low-density boundary condition set by the mass-radius 
relationship of super-Earths provides a simple way to infer the
minimum atmospheric height (MAH) for an exoplanet by using just a precise
measurement of the planet's mass ($M_P$) and radius ($R_P$). This determination
serves to both identify promising targets for follow-up atmospheric
characterization as well as identifying implausible solutions derived from
blind spectral retrieval. In \S\ref{sec:method}, we outline a simple, quantitative 
method to infer the confidence of a small planet having an atmosphere and the 
minimum atmospheric height. In \S\ref{sec:examples}, we apply the technique to 
several examples including GJ~1214b. In \S\ref{sec:discussion}, we discuss the 
potential applications and limitations of our method.

\section{A Statistical Method to Infer the Minimum Atmospheric Height (MAH)}
\label{sec:method}

\subsection{Method}
\label{sub:method}

The simple premise of our method is that any small planet found to have a mass
and radius locating it beyond the boundary condition of a pure-water planet
must maintain an atmosphere. More specifically, if we find 
$R_P>R_{P,\mathrm{H20}}(M_P)$ then the minimum atmospheric height (MAH) is 
given by:

\begin{align}
R_{\mathrm{MAH}}(R_P,M_P) = R_P - R_{P,\mathrm{H20}}(M_P).
\label{eqn:RMAH}
\end{align}

In this expression, $M_P$ and $R_P$ are the observed planetary mass and radius
and the latter should be be interpreted as the radius of the solid/liquid core 
of the planet plus any opaque atmosphere. The $R_{P,\mathrm{H20}}$ term denotes 
the theoretical radius of the planet composed purely from non-gaseous H$_2$O 
(given an observed mass $M_P$) and thus extends from the centre of the planet to 
the solid/liquid surface.

An important caveat is that if a planet does not violate this boundary 
condition, our method says nothing about whether the planet does or does 
not have an atmosphere\footnote{An example of such a case could be the 
unlikely scenario (see e.g. \citealt{rogers+seager2010b}) of a dry rocky core 
surrounded by a thick hydrogen/helium envelope (no water) but with 
$R_P<R_{\mathrm{MAH}}$}.

A typical analysis of a recently discovered super-Earth/mini-Neptune includes
a mass-radius plot showing the various contours for different potential
compositions and a cross marking the position of the new found planet. 
Usually, the width and height of the cross denote the 68.3\% quantile 
confidence range of the planet's mass and radius, respectively. Consider that
a planet resides slightly above the mass-radius contour of a 100\% water 
planet. Using just the $M_P$- and $R_P$-axis error bars, one cannot reasonably
estimate the confidence level of a planet being significantly above this
contour, and thus maintaining an atmosphere. This is because the posterior joint
probability distribution may be (and often is) non-Gaussian, correlated and/or 
multi-modal.

We propose calculating the term $R_{\mathrm{MAH}}$ using realizations of 
$M_P$-$R_P$ drawn from the posterior joint probability distribution of the
system parameters\footnote{Note then that our approach therefore has no a-priori 
preference for a particular composition.}. In doing so, we sample the possible 
parameters consistent with the data in a statistically appropriate way and 
collapse the two-dimensional array into a one-dimensional vector describing the 
quantity of interest.

\subsection{Calculating $R_{P,\mathrm{H20}}(M_P)$}

\begin{figure}
\begin{center}
\includegraphics[width=8.4cm,angle=0,clip=true]{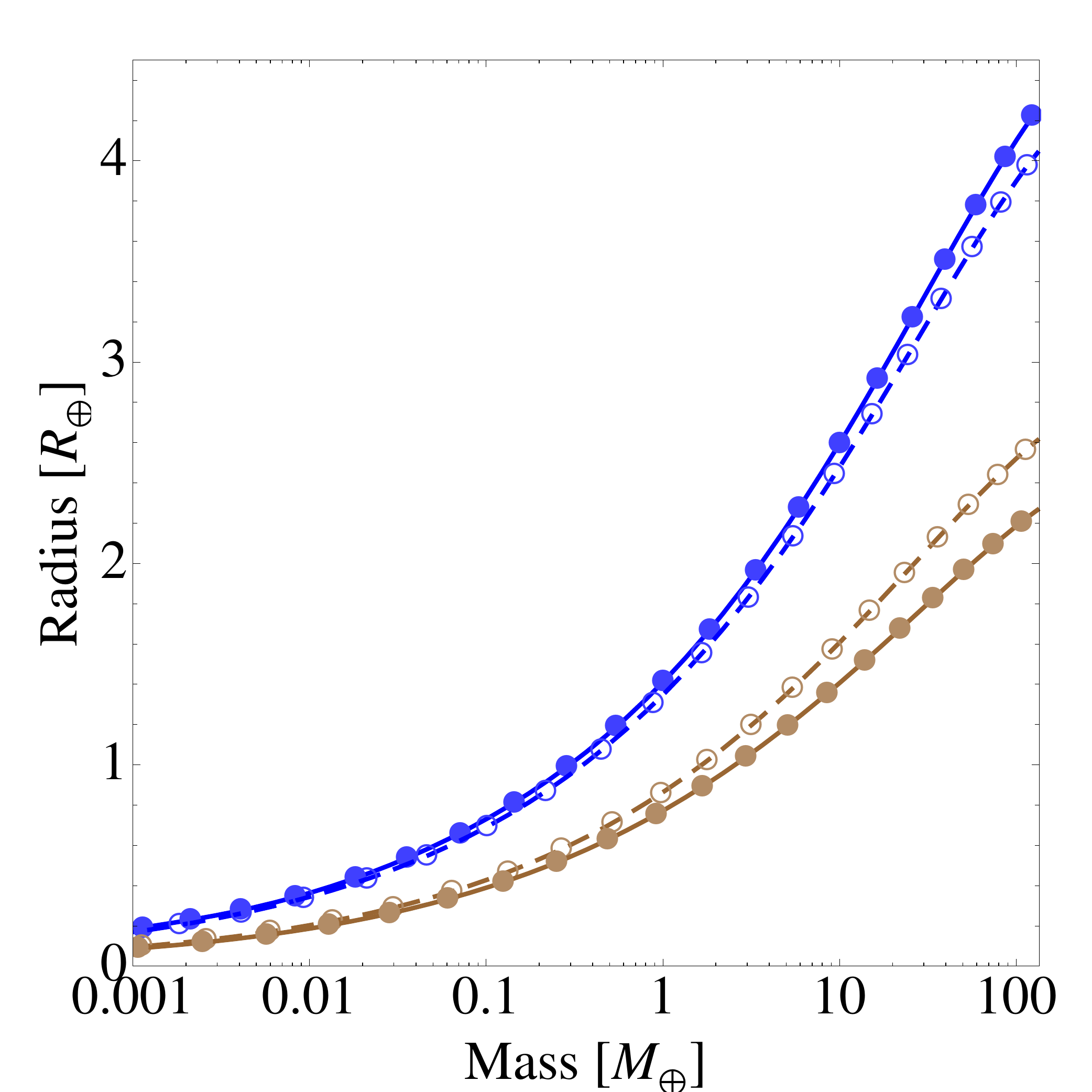}
\caption{\emph{
Mass-radius diagram showing the range of plausible phases for
an atmosphere-less super-Earth (i.e. the boundary conditions), as derived from 
the model of \citet{zeng:2013}.
Points taken from the model are shown as circles, along with our interpolation
line shown overlaid. Blue is that of a 100\%-H$_2$0 planet, blue-dashed 
is 75\%-H$_2$0-25\%-MgSiO$_3$, brown is 100\%-Fe and brown-dashed is 
75\%-Fe-25\%-MgSiO$_3$.}}
\label{fig:zeng}
\end{center}
\end{figure}

In this work, we estimate $R_{P,\mathrm{H20}}(M_P)$ by interpolating the
tables of \citet{zeng:2013}. Fig.~\ref{fig:zeng} shows the 100\%-H$_2$0
mass-radius relation in solid blue. Realizations above this curve
correspond to a positive $R_{\mathrm{MAH}}$. We calculate our interpolation 
by first selecting all $M_P$-$R_P$ entries in the \citet{zeng:2013} catalogue
corresponding to a 100\%-H$_2$0 composition. Plotting $R_P$ as a function of 
$\log_e M_P$ (as shown in Fig.~\ref{fig:zeng}) reveals a smooth behaviour
which may be fitted using a polynomial. We find that a seventh-order polynomial 
describes all of the features well and this function is
supported over the range of masses provided in the \citet{zeng:2013}
catalogue for the 100\%-H$_2$0 composition; specifically 
$4.86\times10^{-4}$\,$M_{\oplus}<M_P<4.86$\,$M_{\oplus}$. The functional
form of our interpolation is given by

\begin{align}
(R_{P,\mathrm{H20}}/R_{\oplus}) = \sum_{n=0}^7 a_n [\log_e (M_P/M_{\oplus})]^n,
\label{eqn:RH20}
\end{align}

where the $a_n$ coefficients are provided in Table~\ref{tab:coeffs}, derived
using a simple least squares regression.

Although the 100\%-H$_2$0 contour is a valid boundary condition, one can
consider it to be somewhat overly conservative. A commonly assumed maximum
initial water content is 75\% (e.g. \citealt{mordasini:2009}) and 
\citet{marcus:2010} have shown that giant impacts cannot increase the 
water fraction. A more realistic boundary condition, then, is to use
75\%-H$_2$0-25\%-MgSiO$_3$. We again find a seventh-order polynomial in 
$\log_e M_P$ well describes the corresponding model
from \citet{zeng:2013}, as shown in Fig.~\ref{fig:zeng} (see
Table~\ref{tab:coeffs} for the corresponding coefficients and 
Table~\ref{tab:support} for the supported range of this model).

\begin{table*}
\caption{\emph{Polynomial coefficients derived for the interpolation functions 
of the super-Earth mass-radius boundary conditions. Each column denotes a 
different composition, where we describe $(R_P/R_{\oplus}) = 
\sum_{n=0}^7 a_n [\log_e (M_P/M_{\oplus})]^n$.}} 
\centering 
\begin{tabular}{c c c c c} 
\hline\hline 
Parameter & 100\%-H$_2$0 & 75\%-H$_2$0-25\%-MgSiO$_3$ & 75\%-Fe-25\%-MgSiO$_3$ & 100\%-Fe \\ [0.5ex] 
\hline 
$a_0$ & 
$+1.409\times10^{+0}$ & 
$+1.346\times10^{+0}$ & 
$+8.633\times10^{-1}$ & 
$+7.714\times10^{-1}$ \\ 
$a_1$ & 
$+3.942\times10^{-1}$ & 
$+3.797\times10^{-1}$ & 
$+2.522\times10^{-1}$ & 
$+2.180\times10^{-1}$ \\ 
$a_2$ & 
$+5.015\times10^{-2}$ & 
$+4.669\times10^{-2}$ & 
$+3.040\times10^{-2}$ & 
$+2.485\times10^{-2}$ \\ 
$a_3$ & 
$+2.513\times10^{-3}$ & 
$+1.992\times10^{-3}$ & 
$+9.476\times10^{-4}$ & 
$+6.916\times10^{-4}$ \\ 
$a_4$ & 
$-4.557\times10^{-4}$ & 
$-3.469\times10^{-4}$ & 
$-2.628\times10^{-4}$ & 
$-2.037\times10^{-4}$ \\ 
$a_5$ & 
$-9.717\times10^{-5}$ & 
$-7.638\times10^{-5}$ & 
$-4.463\times10^{-5}$ & 
$-3.366\times10^{-5}$ \\ 
$a_6$ & 
$-3.900\times10^{-6}$ & 
$-6.315\times10^{-6}$ & 
$-3.001\times10^{-6}$ & 
$-2.271\times10^{-6}$ \\ 
$a_7$ & 
$+1.777\times10^{-7}$ & 
$-1.981\times10^{-7}$ & 
$-7.668\times10^{-8}$ & 
$-5.882\times10^{-8}$ \\ [1ex] 
\hline\hline 
\end{tabular}
\label{tab:coeffs} 
\end{table*}

\begin{table*}
\caption{\emph{Supported range of our interpolation model, based on the extrema
of the models computed by \citet{zeng:2013}.}} 
\centering 
\begin{tabular}{c c c c c} 
\hline\hline 
Parameter & 100\%-H$_2$0 & 75\%-H$_2$0-25\%-MgSiO$_3$ & 75\%-Fe-25\%-MgSiO$_3$ & 100\%-Fe \\ [0.5ex] 
\hline 
$M_{P,\mathrm{min}}$\,$[M_{\oplus}]$ & 
$4.854\times10^{-4}$ & 
$6.197\times10^{-5}$ & 
$6.454\times10^{-6}$ & 
$6.144\times10^{-6}$ \\ 
$M_{P,\mathrm{max}}$\,$[M_{\oplus}]$ & 
$4.864\times10^{+2}$ & 
$3.393\times10^{+2}$ & 
$1.397\times10^{+2}$ & 
$1.321\times10^{+2}$ \\ [1ex] 
\hline\hline 
\end{tabular}
\label{tab:support} 
\end{table*}

\subsection{Confidence of $R_{\mathrm{MAH}}>0$}

Using Equation~\ref{eqn:RMAH}, one may compute the posterior distribution
of $R_{P,\mathrm{H20}}$ and then $R_{\mathrm{MAH}}$. From this latter
distribution, one may easily compute the confidence of the planet in
question maintaining an atmosphere, under the model assumptions. This
is done by evaluating the number of realizations which yield
$R_{\mathrm{MAH}}>0$:

\begin{align}
\mathrm{P}(R_{\mathrm{MAH}}>0) = \frac{\mathrm{\#}\,\,\mathrm{realizations}\,\,\mathrm{where}\,\,R_{\mathrm{MAH}} > 0}{\mathrm{\#}\,\,\mathrm{realizations}\,\,\mathrm{total}}.
\end{align}

\section{Examples}
\label{sec:examples}

\subsection{GJ 1214b}

As a pedagogical example, we consider here perhaps the most well-characterized
small exoplanet to date, GJ 1214b. Originally discovered by 
\citet{charbonneau:2009}, this 2.8\,$R_{\oplus}$ planet orbits a nearby
M4.5 dwarf and consequently there exists a considerable literature of
observations for this system, spanning transits, high resolution stellar
spectra, radial velocities and parallaxes \citep{escude:2012}. The planet
has also been studied extensively using transit spectroscopy in the quest
to identify molecular absorption features (e.g. \citealt{bean:2011,desert:2011,
berta:2012}).

Although the system continues to be intensively observed and thus the physical
parameters of this system are likely to be refined in the near future, we
provide here an applied example of the minimum atmospheric heigh (MAH)
calculation to the planet GJ 1214b. The most recent and comprehensive attempt
to derive self-consistent parameters for the planet and host star comes from
\citet{escude:2012}, who combined an updated trigonometric parallax, medium 
infrared spectroscopy, re-analysed HARPS radial velocities, the photometric 
catalogue and a suite of transit measurements in their analysis.

After obtaining the posterior joint probability distribution of the system 
parameters (personal communication), we computed $R_{\mathrm{MAH}}$ using
Equation~\ref{eqn:RMAH} for a sample of $10^5$ realizations drawn from the
ensemble. For computing $R_{P,\mathrm{H20}}(M_P)$, we assumed the boundary
condition associated with a 75\%-H$_2$0-25\%-MgSiO$_3$ composition from
\citet{zeng:2013}. Table~\ref{tab:examples} provides the results of our
analysis.

\begin{figure}
\begin{center}
\includegraphics[width=8.4cm,angle=0,clip=true]{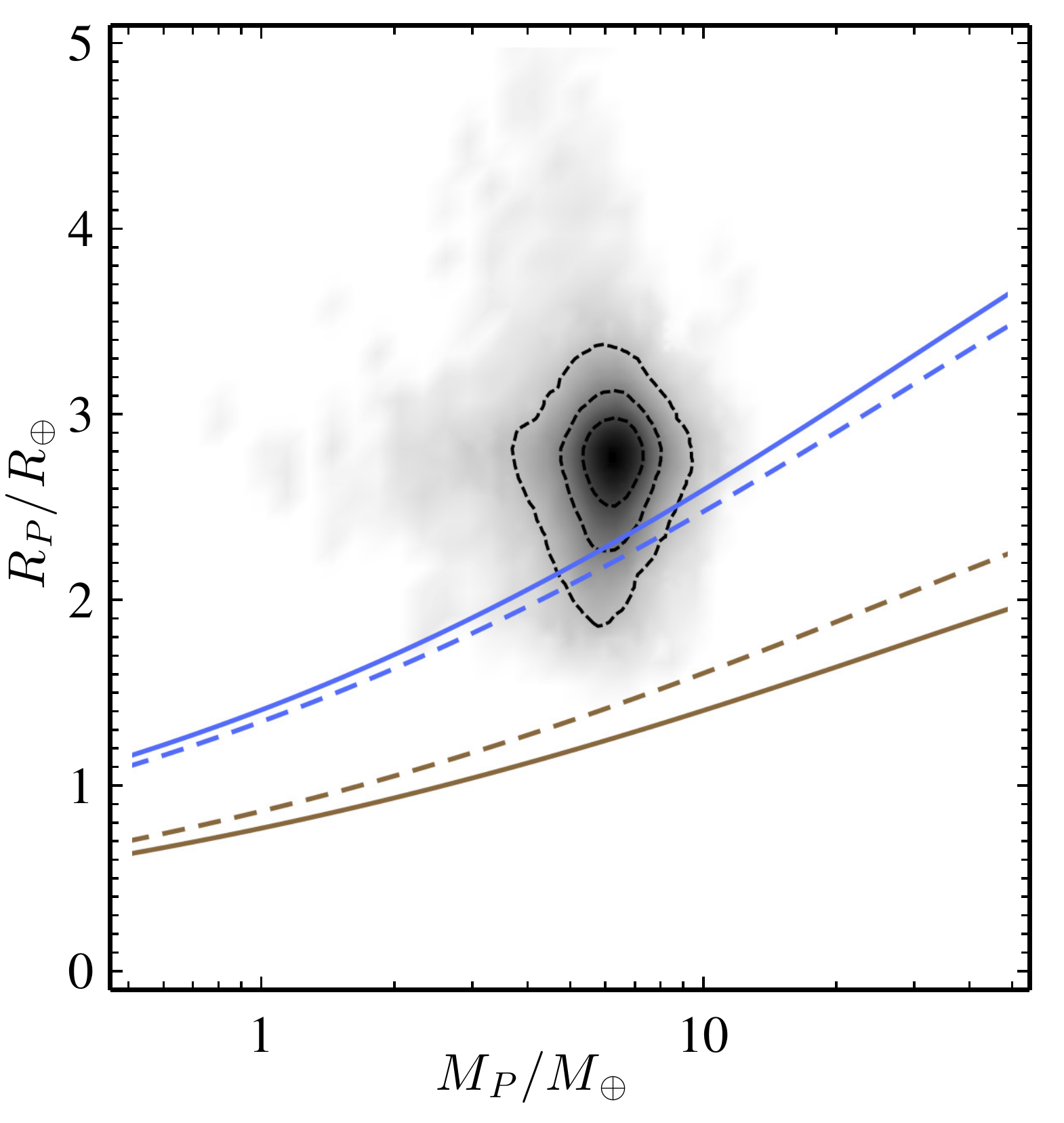}
\caption{\emph{Zoomed-in version of Fig.~\ref{fig:zeng} with the joint 
posterior distribution of GJ~1214b from \citet{escude:2012} shown overlaid.
The three contours correspond to one, two and three-sigma confidence intervals.
97.2\% (2.2\,$\sigma$) of the points lie above the dashed blue line and thus
are inconsistent with an atmosphere-less super-Earth.}}
\label{fig:gj1214}
\end{center}
\end{figure}

\begin{figure}
\begin{center}
\includegraphics[width=8.4cm,angle=0,clip=true]{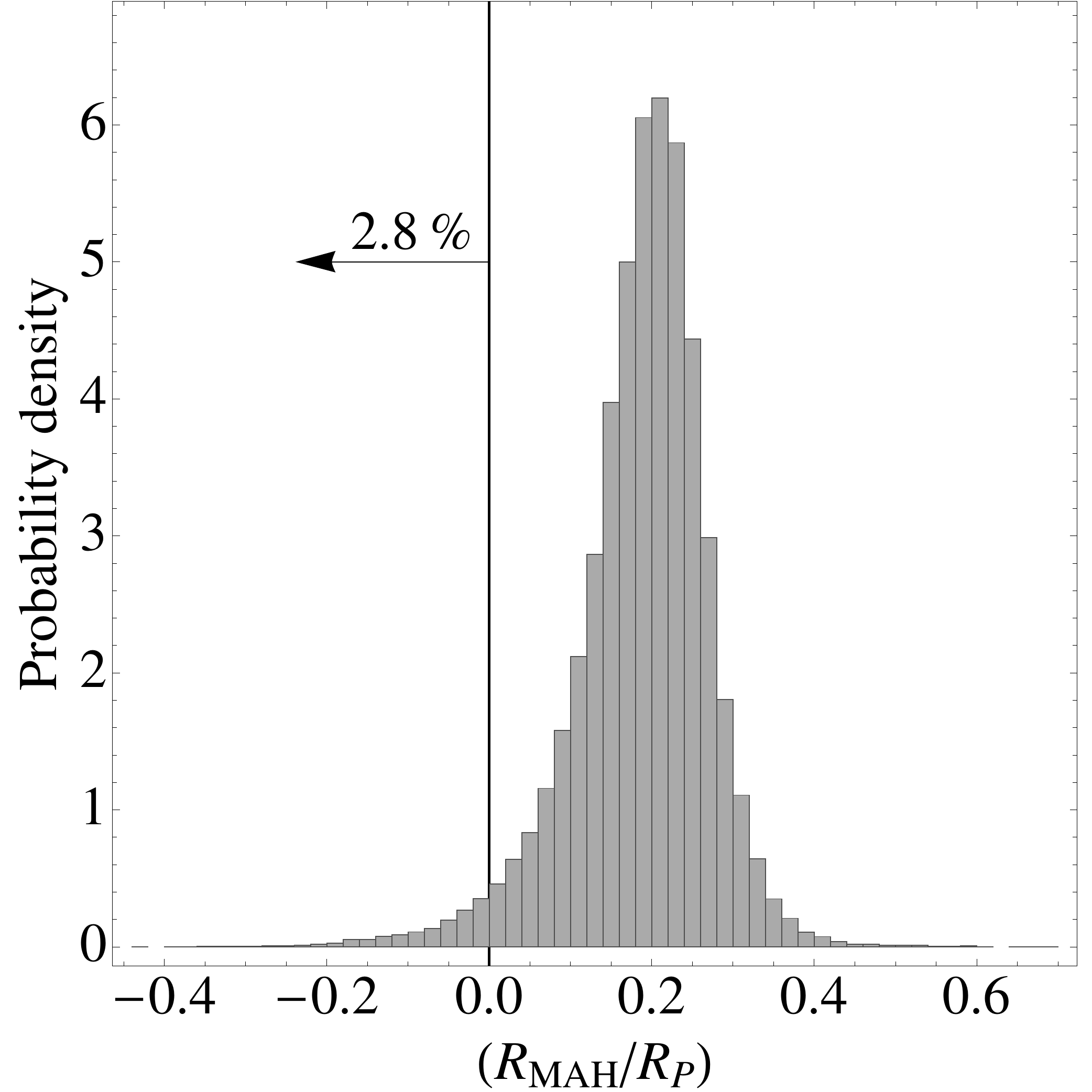}
\caption{\emph{Marginalized posterior distribution of the minimum atmospheric
height ($R_{\mathrm{MAH}}$) normalized by the planetary radius for GJ~1214b. 
Using the joint posteriors of \citet{escude:2012}, the interior structure models 
of \citet{zeng:2013} and Equation~\ref{eqn:RMAH}, we find 
$(R_{\mathrm{MAH}}/R_P)=0.197_{-0.079}^{+0.061}$ and a 97.2\% confidence that
the planet maintains an extended atmosphere.}}
\label{fig:histo}
\end{center}
\end{figure}

We find that 97.2\% (2.2\,$\sigma$) of realizations yield an $M_P$-$R_P$ 
location inconsistent with a super-Earth planet lacking an atmosphere, as shown 
in Fig.~\ref{fig:gj1214} and \ref{fig:histo}. We note that using a 100\%-H$_2$0 
composition boundary condition slightly reduces this to 94.6\%. The minimum 
atmospheric height (MAH) of GJ~1214b is found to be $R_{\mathrm{MAH}} = 
0.54_{-0.24}^{+0.21}$\,$R_{\oplus}$ (quoting the median and $\pm34.1$\% 
quantiles), which translates to $19.7_{-7.9}^{+6.1}$\% of the total planetary 
radius.

\citet{rogers+seager2010b} pointed out that the reported mass and
radius (together with the quoted uncertainties) of GJ~1214b from
\citet{charbonneau:2009} suggest that the object almost surely
has a gas atmosphere layer.  Here, we have quantified the credibility
of this inference by using the full joint posterior probability
density function for mass and radius.  GJ~1214b has an equilibrium
temperature in the vicinity of $\sim$400---550~K, depending on its
Bond albedo.  If the atmosphere is convective and nearly adiabatic
from roughly the transit radius to the base of the gas envelope (at $R
\approx 2.25 R_{\oplus}$), then the adiabatic compression results in
an extremely hot envelope base if the atmosphere is significantly
heavier than an H/He composition.  The observed featureless transit
spectrum (\citealt{bean_et_al2010, bean:2011, berta:2012}
--- although note that \citealt{croll_et_al2011} found that the
transit spectrum was {\it not} featureless) suggests either a high
mean molecular weight atmosphere or, if the composition is H/He, a
high cloud layer masking the features that would otherwise be seen due
to the large scale height \citep{miller-ricci_et_al2009a}.  Our
analysis cautions against drawing any definitive conclusions about
chemical composition from a featureless spectrum for this planet,
given the current joint posterior for mass and radius.  The best-fitting
mass and radius values do suggest (if the spectrum is featureless)
that either clouds in an H/He atmosphere or a small scale height due
to heavy composition (e.g., H$_2$O) prevent observable variations with
wavelength in the transit radius; but, as is clear in Fig. 2,
solutions requiring arbitrarily little atmosphere remain consistent
with the radial velocity and transit data, if the bulk composition is
fairly light (pure water or water+silicate).

\subsection{Other examples}

We briefly comment that Neptune and Uranus both yield a positive
$R_{\mathrm{MAH}}$, which is consistent with the estimate core-sizes of these
worlds (see Table~\ref{tab:examples}).

We also demonstrate the MAH calculation on four other exoplanet examples.
KOI-142b is a planet detected by the transit technique and confirmed through
transit timing variations (TTV) \citep{nesvorny:2013} and is somewhat larger 
than that what might be typically associated with a ``small'' exoplanet. It is
perhaps not surprising then that we find that the planet unambiguously has an 
atmosphere (posteriors obtained through personal communication). 

Similarly, the planet Kepler-36c \citep{carter:2013} has a radius close
to that of Neptune but a lower mass and we find that every single posterior 
sample supports an extended atmosphere using $R_{\mathrm{MAH}}$. However,
the other planet in the system, Kepler-36b, is significantly smaller at 
$\sim1.5$\,$R_{\oplus}$ with a mass of $\sim4.5$\,$M_{\oplus}$. In sharp
contrast to Kepler-36c, we find every single posterior sample yields a negative
$R_{\mathrm{MAH}}$. This planet is therefore an example of where the minimum
atmospheric height method says nothing about whether this world does or does
not have an atmosphere, as discussed earlier in \S\ref{sub:method}.

Finally, we consider Kepler-22b which was detected by \citet{borucki:2012} and 
is similar in radius to GJ~1214b but lies in the habitable-zone of the host 
star. Determining whether the planet has a gaseous envelope or not is therefore 
particularly important due to the potential for habitability. The current radial 
velocity measurements for this system only place an upper limit on the planetary 
mass, limiting our ability to measure $R_{\mathrm{MAH}}$. Using posteriors from 
the recent re-analysis of \citet{kipping:2013}, which includes more transit 
data, we find that planet straddles the boundary condition evenly and is 
consistent with either a water-world or water-world with a gas envelope.

\begin{table*}
\caption{\emph{Example calculations of the minimum atmospheric height (MAH)
for several planets. For Solar System planets we quote the equatorial radius and
assume sphericity, as is done for exoplanets. 
}}
\centering 
\begin{tabular}{c c c c c c c} 
\hline\hline 
Planet & $M_P$\,[$M_{\oplus}]$ & $R_P$\,[$R_{\oplus}]$ & $\rho_P$ [g\,cm$^{-3}$] & $R_{\mathrm{MAH}}$\,[$R_{\oplus}]$ & $(R_{\mathrm{MAH}}/R_P)$ & $\mathrm{P}(R_{\mathrm{MAH}}>0)$ [\%] \\ [0.5ex] 
\hline 
GJ-1214b & $6.19_{-0.80}^{+0.80}$ & $2.75_{-0.24}^{+0.18}$ & $1.66_{-0.38}^{+0.56}$ & $+0.54_{-0.24}^{+0.21}$ & $+0.197_{-0.079}^{+0.061}$ & $97.2$ \\
KOI-142b & $6.6_{-6.1}^{+5.9}$ & $4.23_{-0.39}^{+0.30}$ & $0.48_{-0.45}^{+0.54}$ & $+2.07_{-0.65}^{+1.00}$ & $+0.47_{-0.12}^{+0.26}$ & $>99.9$ \\
Kepler-22b & $6.9_{-6.2}^{+20.9}$ & $2.396_{-0.181}^{+0.088}$ & $2.4_{-2.2}^{+7.5}$ & $+0.11_{-0.87}^{+1.04}$ & $+0.05_{-0.37}^{+0.44}$ & $54.5$ \\
Kepler-36b & $4.46_{-0.27}^{+0.34}$ & $1.487_{-0.035}^{+0.034}$ & $7.47_{-0.59}^{+0.72}$ & $-0.537_{-0.047}^{+0.042}$ & $-0.362_{-0.038}^{+0.034}$ & $<0.01$\\
Kepler-36c & $8.09_{-0.45}^{+0.60}$ & $3.682_{-0.056}^{+0.052}$ & $0.891_{-0.045}^{+0.065}$ & $+1.327_{-0.053}^{+0.049}$ & $+0.361_{-0.012}^{+0.009}$ & $>99.99$ \\
\hline
Neptune & 17.147 & 3.883 & 1.64 & +1.08 & +0.277 & - \\
Uranus  & 14.536 & 4.007 & 1.27 & +2.71 & +0.325 & - \\
Earth   & 1.000  & 1.000 & 5.52 & -0.35 & -0.350 & - \\ [1ex]
\hline\hline 
\end{tabular}
\label{tab:examples} 
\end{table*}

\section{Discussion}
\label{sec:discussion}

In this short paper, we have presented a simple, quantitative method to infer 
the minimum atmospheric height of an atmosphere (MAH) for an exoplanet using 
just a precise (and accurate) mass and radius measurement. In cases where the
the $R_{\mathrm{MAH}}>0$ with high confidence, one infers the presence of an
exoplanet atmosphere without ever taking a spectrum of the planet.

We envision that this metric will aid in the interpretation of exoplanet transit 
spectra by providing an additional boundary condition. For a given interior mass 
and radius ($M_P^{\rm int}$ and $R_P^{\rm int}$), and a given total mass and 
(transit) radius ($M_P$ and $R_P$), there are various possible atmospheric 
compositions. Imposing the constraint that the atmosphere should not be denser 
than the interior implies that some possible chemical compositions of the 
atmosphere are disallowed.  Loosely speaking, one can think of a maximum mean 
molecular weight $\mu_{\rm max}$ for each possible ``boundary condition'' vector
$\mathbf{B} = (M_P^{\rm int},R_P^{\rm int},M_P,R_P,T_{\rm eq})$, where 
$T_{\rm eq}$ is the atmospheric temperature at the transit radius\footnote{In 
actuality, the atmospheric structure depends not just on the mean molecular 
weight but also on the specific chemical composition.}. However, determining 
$\mu_{\rm max}[\mathbf{B}]$ would require a specific equation of state for each 
possible composition. In principle, the information in Fig.~\ref{fig:gj1214},
for the example of GJ~1214b, is sufficient to allow for a calculation of 
$\mu_{\rm max}$, but this exercise is beyond the scope of this paper. But it 
seems clear that, as has been appreciated previously for GJ~1214b, light (H/He) 
and moderately heavy (e.g., H$_2$O) atmospheres are consistent with the data 
but, since the atmosphere appears to be $\sim$20\% the radius of the planet, 
compositions much heavier than H$_2$O are disfavoured.

The simplicity and observationally ``cheap'' nature of determining 
$R_{\mathrm{MAH}}$ makes the metric attractive for general use within the
exoplanet community. However, we do caution here that there are several 
limitations to appreciate when interpreting the minimum atmospheric height. 
Most importantly, the determination is fundamentally a model-dependent one, 
where in this work we used the models of \citet{zeng:2013}. Their model results 
rely on the most recent equations of states and experimental or theoretically 
determined properties of the bulk planetary materials. These will undoubtedly 
improve in the future and some dramatic surprises cannot be excluded, though 
appear unlikely. However, as far as our proposed method is concerned, it is 
trivial to replace this model with whatever mass-radius relation one prefers.

Considering the mass range of super-Earths, the boundary condition of a 
minimum contour of a pure-water planet, might still undergo significant 
correction in the planet models. Apart from the trivial uncertainty due to not 
knowing what is the maximum allowable water fraction (from formation and 
evolution), there is little understanding of the amount of mixing that could 
occur between a water and a H/He envelope (e.g. \citealt{nettelmann:2011}). It 
is likely that this occurs at a particular narrow range of pressures, and hence 
will be dependent on planetary mass, introducing additional structure in the 
mass-radius diagram.

Aside from aiding in the interpretation of exoplanetary spectra, our metric 
provides a quick and cost-effective method to assist in the selection and 
planning of follow-up atmospheric characterization for exoplanets. This is 
particularly important in light of the burgeoning catalogue of exoplanets and 
the upcoming planned missions for exoplanet characterization such as EChO 
\citep{tinetti:2012}. It should be trivial for observers to calculate
$R_{\mathrm{MAH}}$ from their parameter posteriors (using 
Equation~\ref{eqn:RMAH}, Equation~\ref{eqn:RH20} and coefficients from 
Table~\ref{tab:coeffs}) and thus provide a statistically meaningful statement 
regarding the presence of an extended atmosphere, which will surely aid in the 
selection and planning of follow-up observations.

\section*{Acknowledgements}

DMK is supported by the NASA Carl Sagan Fellowships.
DSS gratefully acknowledges support from NSF grant AST-0807444 and the Keck 
Fellowship, and the Friends of the Institute. We are very grateful to 
Guillem Anglada-Escud\'e and collaborators for kindly sharing their posteriors 
with us for GJ~1214b and to Josh Carter for useful conversations on Kepler-36.
Special thanks to the anonymous reviewers for their helpful comments.



%

\bsp

\label{lastpage}


\begin{thebibliography}{99}
\bibitem[\protect\citeauthoryear{Agol et al.}{2010}]{agol:2010} 
Agol, E., Cowan, N. B., Knutson, H. A., Deming, D., Steffen, J. H., 
Henry, G. W. \& Charbonneau, D., 2010, ApJ, 721, 1861
\bibitem[\protect\citeauthoryear{Anglada-Escud\'e et al.}{2013}]{escude:2012} 
Anglada-Escud\'e, G., Rojas-Ayala, b., Boss, A. P., Weinberger, A. J. \& 
Lloyd, J. P., 2013, A\&A, 551, 48
\bibitem[\protect\citeauthoryear{Batalha et al.}{2013}]{batalha:2013} 
Batalha, N. M. et al., 2013, ApJS, 204, 24
\bibitem[{{Bean} {et~al.}(2010){Bean}, {Miller-Ricci Kempton}, \&
  {Homeier}}]{bean_et_al2010}
{Bean}, J.~L., {Miller-Ricci Kempton}, E., \& {Homeier}, D. 2010, Nature, 468,
  669
\bibitem[\protect\citeauthoryear{Bean et al.}{2011}]{bean:2011} 
Bean, J. L. et al., 2011, ApJ, 743, 92
\bibitem[\protect\citeauthoryear{Benneke \& Seager}{2012}]{benneke:2012} 
Benneke, B. \& Seager, S., 2012, ApJ, 753, 100
\bibitem[\protect\citeauthoryear{Berta et al.}{2012}]{berta:2012} 
Berta, Z. K. et al., 2012, ApJ, 747, 35
\bibitem[\protect\citeauthoryear{Borucki et al.}{2012}]{borucki:2012} 
Borucki, W. J. et al., ApJ, 745, 120
\bibitem[\protect\citeauthoryear{Carter et al.}{2012}]{carter:2013} 
Carter, J. A. et al., 2012, Science, 337, 667
\bibitem[\protect\citeauthoryear{Charbonneau et al.}{2002}]{charbonneau:2002} 
Charbonneau, D., Brown, T. M., Noyes, R. W. \& Gilliland, R. L.,
2002, ApJ, 568, 377
\bibitem[\protect\citeauthoryear{Charbonneau et al.}{2009}]{charbonneau:2009} 
Charbonneau, D. et al., 2009, Nature, 462, 891
\bibitem[{{Croll} {et~al.}(2011){Croll}, {Albert}, {Jayawardhana},
  {Miller-Ricci Kempton}, {Fortney}, {Murray}, \& {Neilson}}]{croll_et_al2011}
{Croll}, B., {Albert}, L., {Jayawardhana}, R., {Miller-Ricci Kempton}, E.,
  {Fortney}, J.~J., {Murray}, N., \& {Neilson}, H. 2011, ApJ, 736, 78
\bibitem[\protect\citeauthoryear{D\'esert et al.}{2011}]{desert:2011} 
D\'esert, J.-M. et al., 2011, ApJ, 731, 40
\bibitem[\protect\citeauthoryear{Dong \& Zhu}{2012}]{dong:2012} 
Dong, S. \& Zhu, ApJ, submitted (astro-ph/1212.4853)
\bibitem[\protect\citeauthoryear{Fraine et al.}{2013}]{fraine:2013} 
Fraine, J. D. et al., 2013, ApJ, 765, 127
\bibitem[\protect\citeauthoryear{Fressin et al.}{2013}]{fressin:2013} 
Fressin, F. et al., 2013, ApJ, 766, 81
\bibitem[\protect\citeauthoryear{Kipping et al.}{2013}]{kipping:2013} 
Kipping, D. et al., 2013, ApJ, submitted (astro-ph/1306.1530)
\bibitem[\protect\citeauthoryear{Marcus et al.}{2010}]{marcus:2010} 
Marcus, R. A., Sasselov, D., Stewart, S. \& Hernquist, L., 2010,
ApJ, 719, 45
\bibitem[{{Miller-Ricci} {et~al.}(2009){Miller-Ricci}, {Seager}, \&
  {Sasselov}}]{miller-ricci_et_al2009a}
{Miller-Ricci}, E., {Seager}, S., \& {Sasselov}, D. 2009, ApJ, 690, 1056
\bibitem[\protect\citeauthoryear{Mordasini et al.}{2009}]{mordasini:2009} 
Mordasini, C., Alibert, Y. \& Benz, W., 2009, A\&A, 501, 1139
\bibitem[\protect\citeauthoryear{Nesvorny et al.}{2013}]{nesvorny:2013} 
Nesvorny, D. et al., 2013, ApJ, submitted (astro-ph/1304.4283)
\bibitem[\protect\citeauthoryear{Nettelmann et al.}{2011}]{nettelmann:2011} 
Nettelmann, N., Fortney, J. J., Kramm, U. \& Redmer, R., 2011, ApJ, 733, 2
\bibitem[\protect\citeauthoryear{Pollack et al.}{1996}]{pollack:1996} 
Pollack, J. B., Hubickyj, O., Bodenheimer, P., Lissauer, J. J., Podolak, 
M., Greenzweig, Y., 1996, Icarus, 124, 62
\bibitem[\protect\citeauthoryear{Schneider et al.}{2011}]{schneider:2011} 
Schneider J., Dedieu C., Le Sidaner P., Savalle R., Zolotukhin I., 2011, A\&A, 
532, A79
\bibitem[{{Rogers} \& {Seager}(2010)}]{rogers+seager2010b}
{Rogers}, L.~A. \& {Seager}, S. 2010, ApJ, 716, 1208
\bibitem[\protect\citeauthoryear{Seager \& Deming}{2010}]{seager:2010} 
Seager, S. \& Deming, D., 2010, A\&A, 48, 631
\bibitem[\protect\citeauthoryear{Sing et al.}{2012}]{sing:2012} 
Sing, D. K. et al., 2012, MNRAS, 426, 1663
\bibitem[\protect\citeauthoryear{Tinetti et al.}{2007}]{tinetti:2007} 
Tinetti, G. et al., 2007, Nature, 448, 169
\bibitem[\protect\citeauthoryear{Tinetti \& Beaulieu}{2009}]{tinetti:2009} 
Tinetti, G. \& Beaulieu, J. P., 2009, Proc. IAU Symposium No. 253, Queloz, D.,
Sasselov, D., Torres, M., \& Holman, M., eds., p. 231
\bibitem[\protect\citeauthoryear{Tinetti et al.}{2012}]{tinetti:2012} 
Tinetti, G. et al., 2012, Exp. Astron., 34, 311
\bibitem[\protect\citeauthoryear{Valencia et al.}{2006}]{valencia:2006} 
Valencia, D., O'Connell, R. \& Sasselov, D., 2006, Icarus, 181, 545
\bibitem[\protect\citeauthoryear{Zeng \& Sasselov}{2013}]{zeng:2013} 
Zeng, L. \& Sasselov, D., 2013, PASP, 125, 227
\bibitem[Schneider et al.(2011)]{schneider:2011} Schneider, J.,
  Dedieu, C., Le Sidaner, P., Savalle, R., \& Zolotukhin, I.\ 2011,
  A\&A, 532, A79
\end{thebibliography}
\end{document}